\documentclass{article}
\usepackage[letterpaper,top=1in,bottom=1in,left=1in,right=1in,marginparwidth=0.5in]{geometry}
\usepackage[backend=biber, style=alphabetic, doi=true, url=true, eprint=true]{biblatex}
\usepackage{multicol}
\usepackage{lscape}
\newboolean{preprint}
\setboolean{preprint}{false}
\usepackage[utf8]{inputenc}
\PassOptionsToPackage{hyphens}{url}\usepackage{hyperref}
\usepackage{amsmath}
\usepackage{amsthm}
\usepackage{enumitem}
\usepackage{amssymb}
\usepackage{csquotes}
\usepackage[english]{babel}
\usepackage{braket}
\usepackage{pifont} 
\usepackage{comment}
\usepackage{booktabs}
\usepackage{tabularx}
\usepackage{wasysym}
\usepackage{ragged2e}
\usepackage{caption}
\usepackage{subcaption}
\usepackage{array}
\usepackage{authblk}
\usepackage{bm}
\usepackage[linesnumbered,ruled,vlined]{algorithm2e}
\usepackage{svg}
\usepackage{tikz}
\usepackage{tikz-cd}
\usepackage{mathtools}
\usetikzlibrary{shapes.geometric, intersections, through}
\usetikzlibrary{matrix,positioning,decorations.pathreplacing}
\usetikzlibrary{calc}

\definecolor{myyellow}{rgb}{0.788, 0.62, 0.063}

\DeclarePairedDelimiter\round{\lfloor}{\rceil}

\makeatletter
\def\bb{bb}
\@tfor\next:=ABCDEFGHIJKLMNOPQRSTUVWXYZ\do
 {\begingroup\edef\x{\endgroup
    \noexpand\@namedef{\bb\next}{\noexpand\mathbb{\next}}%
  }\x}
\makeatother

\makeatletter
\def\cal{cal}
\@tfor\next:=ABCDEFGHIJKLMNOPQRSTUVWXYZ\do
 {\begingroup\edef\x{\endgroup
    \noexpand\@namedef{\cal\next}{\noexpand\mathcal{\next}}%
  }\x}
\makeatother

\newcolumntype{L}[1]{>{\hsize=#1\hsize\RaggedRight} X}

\newif\ifphysics
\physicstrue
\newif\ifmath
\mathfalse

\ifphysics
    
\fi
\ifmath
    
\fi



\begin{document}

%TC:ignore
\title{Real-time decoding of the gross code memory with FPGAs.}
\ifthenelse{\boolean{preprint}}{
\author{Thilo Maurer}
\email[Contact: ]{tmaurer@de.ibm.com}
\affiliation{IBM Quantum}
\author{Markus~Bühler}
\affiliation{IBM Quantum}
\author{Michael Kr\"oner}
\affiliation{IBM Quantum}
\author{Frank Haverkamp}
\affiliation{IBM Quantum}
\author{Tristan Müller}
\affiliation{IBM Quantum}
\author{Drew Vandeth}
\affiliation{IBM Quantum}
\author{Blake R. Johnson}
\affiliation{IBM Quantum}
}{
\author{Thilo Maurer\thanks{Contact: tmaurer@de.ibm.com}}
%\email[Contact: ]{tmaurer@de.ibm.com}
\author[1]{Markus~Bühler}
\author[1]{Michael Kr\"oner}
\author[1]{Frank Haverkamp}
\author[1]{Tristan Müller}
\author[1]{Drew Vandeth}
\author[1]{Blake R. Johnson}
\affil[1]{IBM Quantum}

}

\date{Oct 24, 2025} 

\ifthenelse{\boolean{preprint}}{}{\maketitle}

\begin{abstract}
    We introduce a prototype FPGA decoder implementing the recently discovered Relay-BP algorithm and targeting memory experiments on the $[[144,12,12]]$ bivariate bicycle quantum low-density parity check code. The decoder is both fast and accurate, achieving a belief propagation iteration time of 24ns. It matches the logical error performance of a floating-point implementation despite using reduced precision arithmetic. This speed is sufficient for an average per cycle decoding time under $1\,\mathrm{\mu s}$ assuming circuit model error probabilities are less than $3 \times 10^{-3}$. This prototype decoder offers useful insights on the path toward decoding solutions for scalable fault-tolerant quantum computers.
\end{abstract}

\ifthenelse{\boolean{preprint}}{\maketitle}{}
%TC:endignore

\section{Introduction}

The pursuit of fault-tolerant quantum computing promises to unlock a wide array of transformative applications. Recent advances in fault-tolerant architectures~\cite{tourdegross} have positioned quantum low-density parity check (LDPC) codes as a leading approach, with the bicycle architecture emerging as a strong candidate for realizing the first practical fault-tolerant quantum systems~\cite{ibmroadmap}. However, a key challenge remains: the development of a viable, real-time quantum error correction (QEC) decoder tailored to quantum LDPC codes.

Addressing this challenge involves a two-step process. First, it is essential to design a decoding algorithm that is accurate, flexible, and amenable to efficient hardware implementation~\cite{muller2025}. Second, the algorithm must be physically realized in hardware in a way that meets the stringent performance demands of fault-tolerant quantum computing. Specifically, any practical decoder must be compact and fast---where “fast” encompasses both high throughput and low latency, ensuring corrections are computed faster than new syndrome data arrives in each QEC cycle.~\cite{Terhal2015, Holmes-et-al-2020}.

Earlier this year, the Relay-BP quantum LDPC decoder was proposed~\cite{muller2025}, offering a promising solution to the first step. It achieves strong accuracy and flexibility while being well-suited for hardware implementation, positioning its physical realization as the next major milestone in decoder development.

The need for high throughput and low latency is especially pronounced in superconducting qubit architectures, where QEC cycles operate on microsecond timescales. In such systems, decoders must have ultra-low latencies. Throughput and latency are tightly coupled to the underlying technology choices. While GPU-based and AI-accelerator platforms~\cite{alphaqubit} offer programmability and potential throughput advantages, they suffer from significant latency overheads due to data transfer and kernel launch delays---posing a challenge for real-time decoding.

As a result, for both surface code decoders and quantum LDPC codes, many researchers have turned to Field Programmable Gate Arrays (FPGAs)~\cite{Wu_2025, Liyanage_2024, caune2024demonstratingrealtimelowlatencyquantum, ziad2024localclusteringdecoderfast, Liao-et-al-2023, Valls-et-al-2021, maan2025} or Application Specific Integrated Circuits (ASICs)~\cite{barber-et-al-2025, Kadomoto-et-al-2025, Bascones-2025, Das-et-al-2022}. While ASICs offer excellent performance characteristics, they are very costly to develop and less flexible. In contrast, FPGAs provide a compelling balance of performance and cost, with more straightforward setup and integration into a quantum control system --- making them a practical choice for the next phase of decoder prototypes.

In this manuscript, we present an FPGA implementation of the Relay-BP decoding algorithm~\cite{muller2025}, applied to memory experiments on bivariate bicycle codes~\cite{bravyi2023, tourdegross}, including the  Loon [[56,2,10]] and gross [[144,12,12]] codes. Building upon the work of Valls et al.~\cite{Valls-et-al-2021}, our implementation explores the speed limits of Relay-BP in a maximally parallel computational architecture. This approach trades space for time, making it resource-intensive and pushing the boundaries of what can be achieved on a single modern FPGA.

A factor that strongly influences the resource utilization is the numeric representation of the messages. Consequently, our study examines the sensitivity of Relay-BP decoding accuracy to reduced precision arithmetic, finding that for split X/Z-decoding as few as 4-bits of precision are sufficient to match a floating point implementation,  while 6 bits are sufficient for correlated XYZ-decoding. We find that an \texttt{int4} Relay-BP split X/Z decoder of a 12+1 cycle syndrome cycle on the gross code may be realized in a AMD VU19P FPGA and achieve a Relay-BP iteration time of 24ns. Under an academic circuit noise model with $p < 10^{-3}$, we find that Relay-BP-1 converges in less than 10 iterations on average. Under these conditions, our FPGA solution would decode 12-cycle windows of repeated syndrome cycles in less than 240ns on average. This is faster than our anticipated syndrome collection rate of $1\,\mathrm{\mu s}$ per cycle, implying that the FPGA decoder has sufficiently low latency to avoid the backlog problem~\cite{Terhal2015}.

% \blake{Perhaps save this paragraph for v2?} The maximally-parallel implementation introduces scalability challenges to apply this approach to larger codes such as the merged codes encountered in logical operations~\cite{tourdegross}. Thus, we also describe a method to exploit temporal symmetry in decoding windows of repeated cycles to cut the resource utilization by large integer factors.

This paper is organized as follows: we first review the essential elements of decoding in order to introduce relevant notation. We then provide a summary of the Relay-BP algorithm implemented in the decoder. We introduce the architecture of our FPGA implementation, examine its resource utilization and our validation methods. Finally, we examine the performance of the decoder and provide remarks on possible next steps.
\section{Quantum Error Correction Decoding Framework}

%Classical and quantum error-correcting codes share core principles but differ significantly due to the nature of the information they protect. In the classical case, errors affect bits in discrete ways as bit flips. These errors can be detected using syndrome data, which provides indirect clues about the error without revealing its exact form. Decoding involves using this data to infer and apply a correction. In the quantum case, errors form a continuum due to the analog nature of quantum states. Initially, this led to skepticism about whether quantum error correction was even possible. However, quantum codes were developed that discretize errors into bit flips and phase flips, and combinations thereof — and allow for the indirect extraction of syndrome data without collapsing the quantum state. Quantum codes also introduce degeneracy, a unique feature where different error patterns can produce the same syndrome. This complicates decoding, as multiple distinct errors may be equally valid corrections, requiring more sophisticated inference strategies.

%\subsection{Decoding Framework}

A unified language exists~\cite{ott2025decisiontreedecodersgeneralquantum} that describes both classical and QEC decoding in terms of linear algebra, which we will utilize here. We begin by restating the definition of a decoder in that terminology:

\paragraph{Errors and decoding}
Consider a check matrix $\mathbf{H} \in \mathbb{F}_2^{M\times N}$, a logical action matrix $\mathbf{A} \in \mathbb{F}_2^{K\times N}$, and a probability vector $\mathbf{p} \in (0, 1/2)^N$. Errors are described by a vector $\mathbf{e} \in \mathbb{F}_2^N$ that we assume to be randomly drawn according to the distribution
\[
\mathrm{Pr}_{\mathbf{p}}(\mathbf{e}) =  \prod_{j =0}^{N-1} (1-p_j) \left(\frac{p_j}{1-p_j}\right)^{e_j}.
\]
\noindent An error $\mathbf{e}$ is related to an observed syndrome $\bm{\sigma} \in \mathbb{F}_2^M$ by the check matrix with $\bm{\sigma} = \mathbf{H}\mathbf{e}$. A decoder is a classical algorithm which,
given $\bm{\sigma}$ (and with knowledge of the objects $\mathbf{H}$, $\mathbf{A}$ and $\mathbf{p}$), proposes a correction $\mathbf{\hat{e}} \in \mathbb{F}_2^N$. We say that the decoder succeeds if both $\mathbf{H}\mathbf{\hat{e}}  = \bm{\sigma}$ and $\mathbf{A}\mathbf{\hat{e}} = \mathbf{A}\mathbf{e}$, and that it fails otherwise.

%\paragraph{Example} Different choices of $\mathbf{H}$, $\mathbf{A}$, and $\mathbf{w}$ define decoding problems across a wide range of settings: from classical codes to separate $X$ and $Z$ de-
%coding for quantum CSS codes~\cite{steane1996multiple,calderbank1996good}, and further to fault-tolerant quantum circuits under correlated circuit noise. For example a classical linear code is defined by its check matrix $\mathbf{H} \in \mathbb{F}_2^{M\times N}$. Here codewords are bitstrings $\mathbf{c} \in \mathbb{F}_2^N$ such that $\mathbf{H}\mathbf{c} = 0$. Then, given a codeword $\mathbf{c}$, an additional error $\mathbf{e}$ results in $(\mathbf{c} + \mathbf{e})$,
%which has syndrome $\bm{\sigma} = \mathbf{H}(\mathbf{c} + \mathbf{e}) = \mathbf{H}\mathbf{e}$. The logical action matrix $\mathbf{A}$ in the case is the identity matrix and
%does not play a role. This means that in the typical scenario of classical codes, all codewords are considered in-equivalent. This is not true in general for quantum codes.

\paragraph{Weights} Given a weight vector $\mathbf{w} \in (0, \infty)^N$, we define the weight of an error vector $\mathbf{e}$ as $w(\mathbf{e}) = \sum_{j=1}^N \omega_je_j$. We use the common weight system called the log likelihood ratio (LLR) which is defined by $\omega_j = \log\frac{1-p_j}{p_j}$ and gives an equivalent representation for the errors.

\paragraph{Graphical representation of check matrices} For a given decoder problem, it is often useful to interpret the check matrix as a graph. For a visual representation, we let ($\ocircle$) denote \emph{error nodes} that correspond to columns of $\mathbf{H}$ and ($\square$) denote \emph{check nodes} that correspond to rows of $\mathbf{H}$. Nodes are connected by an edge if there is a $1$ in the corresponding location of the $\mathbf{H}$ matrix, which implies the resulting decoding graph, $G$, is bipartite. We denote the neighbors of node $n$ in $G$ as $\mathcal{N}(n)$.

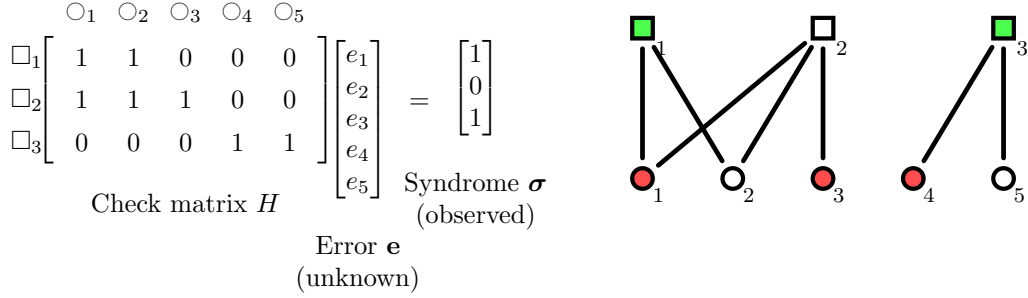
\begin{figure}[h!]

% --- Measure glyph sizes; use them to size drawn nodes exactly ---
\newsavebox{\sqbox}\sbox{\sqbox}{$\square$}
\newlength{\sqW}\settowidth{\sqW}{\usebox{\sqbox}}
\newlength{\sqH}\settoheight{\sqH}{\usebox{\sqbox}}
\newlength{\sqD}\settodepth{\sqD}{\usebox{\sqbox}}
\newlength{\sqSize}\setlength{\sqSize}{\sqH}\addtolength{\sqSize}{\sqD}
\newlength{\sqTmp}\setlength{\sqTmp}{\sqW}
\ifdim\sqTmp>\sqSize \setlength{\sqSize}{\sqTmp}\fi
\newlength{\sqHalf}\setlength{\sqHalf}{.5\sqSize}

\newsavebox{\cibox}\sbox{\cibox}{$\ocircle$}
\newlength{\ciW}\settowidth{\ciW}{\usebox{\cibox}}
\newlength{\ciH}\settoheight{\ciH}{\usebox{\cibox}}
\newlength{\ciD}\settodepth{\ciD}{\usebox{\cibox}}
\newlength{\ciSize}\setlength{\ciSize}{\ciH}\addtolength{\ciSize}{\ciD}
\newlength{\ciTmp}\setlength{\ciTmp}{\ciW}
\ifdim\ciTmp>\ciSize \setlength{\ciSize}{\ciTmp}\fi
\newlength{\ciHalf}\setlength{\ciHalf}{.5\ciSize}

% offsets so subscripts look like true subscripts near node corners
\newlength{\subx}\setlength{\subx}{0.12\sqSize}
\newlength{\suby}\setlength{\suby}{0.12\sqSize}

% ---- control the gap between e-vector and RHS vector ----
\newlength{\rhssep}\setlength{\rhssep}{6mm} % increase/decrease as needed
\centering
\begin{tikzpicture}[baseline={(current bounding box.center)}]

  %================ Rectangular matrix (with labels) ================%
  \matrix (A) [matrix of math nodes,
               left delimiter={[}, right delimiter={]},
               row sep=3pt, column sep=8pt]
  {
    1 & 1 & 0 & 0 & 0 \\
    1 & 1 & 1 & 0 & 0 \\
    0 & 0 & 0 & 1 & 1 \\
  };

  % Column labels on top (glyphs)
  \foreach \j in {1,...,5} {
    \node[above=3pt of A-1-\j] {$\ocircle_{\j}$};
  }
  % Row labels on left (glyphs)
  \foreach \i in {1,...,3} {
    \node[left=5pt of A-\i-1] {$\square_{\i}$};
  }

  %================ Matrix → e-vector → "=" → RHS vector (tops aligned) ===============%
  % e-vector: top bracket aligned to the matrix top
  \node[anchor=north west] (evec) at ($(A.north east)+(0.6mm,0)$)
    {$\begin{bmatrix} e_1\\ e_2\\ e_3\\ e_4\\ e_5 \end{bmatrix}$};

  % RHS vector: moved further right by \rhssep (still top-aligned via evec)
  \node[anchor=north west] (rhs) at ($(evec.north east)+(\rhssep,0)$)
    {$\begin{bmatrix} 1\\ 0\\ 1 \end{bmatrix}$};

  % "=" EXACTLY centered between vectors (both horizontally & vertically)
  \coordinate (midEW) at ($(evec.east)!0.5!(rhs.west)$);
  \coordinate (midCC) at ($(evec.center)!0.5!(rhs.center)$);
  \node[anchor=center, inner sep=0, outer sep=0] (eq) at ($(midEW |- midCC)$) {$=$};

  %----------- Labels under each block (centered) -----------
  \node[anchor=north, align=center] at ($(A.south)+(0,-2.0mm)$)
    {Check matrix $H$};

  \node[anchor=north, align=center] at ($(evec.south)+(0,-2.0mm)$)
    {\begin{tabular}{c} Error $\mathbf{e}$ \\ (unknown) \end{tabular}};

  \node[anchor=north, align=center] at ($(rhs.south)+(0,-2.0mm)$)
    {\begin{tabular}{c} Syndrome $\bm{\sigma}$ \\ (observed) \end{tabular}};

  % Center of the product block (for graph alignment)
  \coordinate (prodcenter) at ($(evec.center)!0.5!(rhs.center)$);

  %====================  Bipartite graph (right) =====================%
  \tikzset{
    sqnode/.style={draw,rectangle,minimum width=\sqSize,minimum height=\sqSize,
                   inner sep=0pt,outer sep=0pt,line width=1.6pt},
    cirnode/.style={draw,circle,minimum size=\ciSize,
                    inner sep=0pt,outer sep=0pt,line width=1.6pt},
    idx/.style={font=\footnotesize, inner sep=0pt}, % larger subscripts
    edge/.style={line width=1.8pt, line cap=round}
  }

  % vertical half-gap between square and circle rows
  \def\h{1.0}

  % Place graph center ≥ \qquad from product center (use 30mm here)
  \begin{scope}[shift={($(prodcenter)+(30mm,0)$)}]
    % --- Squares (top row) ---
    \node[sqnode,fill=green!70!white] (s1) at (0,\h)   {};
    \node[sqnode]                     (s2) at (2.4,\h) {};
    \node[sqnode,fill=green!70!white] (s3) at (4.8,\h) {};
    \node[idx,anchor=north west] at ([xshift=\subx,yshift=-\suby]s1.south east) {$1$};
    \node[idx,anchor=north west] at ([xshift=\subx,yshift=-\suby]s2.south east) {$2$};
    \node[idx,anchor=north west] at ([xshift=\subx,yshift=-\suby]s3.south east) {$3$};

    % --- Circles (bottom row) ---
    \node[cirnode,fill=red!70!white]  (c1) at (0,-\h)   {};
    \node[cirnode]                    (c2) at (1.2,-\h) {};
    \node[cirnode,fill=red!70!white]  (c3) at (2.4,-\h) {};
    \node[cirnode,fill=red!70!white]  (c4) at (3.6,-\h) {};
    \node[cirnode]                    (c5) at (4.8,-\h) {};
    \node[idx,anchor=north west] at ([xshift=\subx,yshift=-\suby]c1.south east) {$1$};
    \node[idx,anchor=north west] at ([xshift=\subx,yshift=-\suby]c2.south east) {$2$};
    \node[idx,anchor=north west] at ([xshift=\subx,yshift=-\suby]c3.south east) {$3$};
    \node[idx,anchor=north west] at ([xshift=\subx,yshift=-\suby]c4.south east) {$4$};
    \node[idx,anchor=north west] at ([xshift=\subx,yshift=-\suby]c5.south east) {$5$};

    % --- Edges: border-to-border (no gaps)
    \draw[edge,shorten >=\sqHalf, shorten <=\ciHalf] (s1)--(c1);
    \draw[edge,shorten >=\sqHalf, shorten <=\ciHalf] (s1)--(c2);

    \draw[edge,shorten >=\sqHalf, shorten <=\ciHalf] (s2)--(c1);
    \draw[edge,shorten >=\sqHalf, shorten <=\ciHalf] (s2)--(c2);
    \draw[edge,shorten >=\sqHalf, shorten <=\ciHalf] (s2)--(c3);

    \draw[edge,shorten >=\sqHalf, shorten <=\ciHalf] (s3)--(c4);
    \draw[edge,shorten >=\sqHalf, shorten <=\ciHalf] (s3)--(c5);
  \end{scope}
\end{tikzpicture}

    \caption{\label{fig:decoding-basics} The decoding graph visually represents the check matrix: circular ($\ocircle$) \emph{error nodes}  correspond to columns of $\mathbf{H}$, and square ($\square$) \emph{check nodes}  to its rows. 
    Filled or unfilled nodes denote a corresponding value of 1 or  0 in either $\bm{\sigma}$ or $\mathbf{e}$. 
    The decoder relies solely on the syndrome (i.e., the values of the check nodes) to infer a candidate error. 
    Check node $i$ has syndrome $\sigma_i = 1$ if it touches an odd number of error nodes with value one.}
\end{figure}

\paragraph{Decoding strategies} As in~\cite{muller2025} we consider two decoding strategies. Our primary strategy for this paper is \emph{XZ-decoding} which decomposes $\bm{\sigma}$ into $\bm{\sigma}_X$ and $\bm{\sigma}_Z$ which are independently decoded using the derived check matrices $\mathbf{H}_X$ and $\mathbf{H}_Z$ to obtain $\mathbf{\hat{e}}_X$ and $\mathbf{\hat{e}}_Z$. These partial corrections are then combined to infer $\mathbf{\hat{e}}$. 
XZ-decoding simplifies decoding by using smaller objects, but may degrade performance by modeling X and Z errors as independent. We also discuss the more complex decoding strategy called \emph{XYZ-decoding} which directly computes a correction $\mathbf{\hat{e}}$ from the full $\mathbf{H}$ and $\bm{\sigma}$.

We assume a linear circuit-level noise model in which each quantum operation fails independently, with errors sampled from a discrete set of error modes. 
Each single-qubit unitary is followed by a Pauli $X$, $Y$, or $Z$ error, each with probability $p/3$, and each two-qubit unitary by one of the 15 non-identity two-qubit Paulis with probability $p/15$.  State preparations and measurements fail with probability $p$, modeled respectively by orthogonal state preparation or measurement outcome flipping. 
We refer to each such error mode as an error, which corresponds to a column in $\mathbf{H}$ and $\mathbf{A}$, with its probability encoded in the associated entry of the vector $\mathbf{p}$.

\section{Relay-BP}
For completeness, we present a compact description of the Relay-BP algorithm as introduced in~\cite{muller2025}. Relay-BP relies upon a subroutine called DMem-BP that incorporates disordered memory strength distributions into a standard belief propagation (BP) algorithm. Relay-BP builds upon DMem-BP by applying a relay ensembling technique and extending the framework to allow negative memory strengths, resulting in improved decoding performance.

\paragraph{DMem-BP} The DMem-BP algorithm operates by iteratively passing real-valued messages along the edges of a decoding graph $G$. Messages exchanged between check node $i$ and error node $j$ are denoted $\mu_{i \rightarrow j}$ and $\nu_{j \rightarrow i}$, respectively. Each message conveys a local log-likelihood belief about the presence of an error at its node: negative values indicate a higher likelihood of error, while positive values suggest no error.

The algorithm is defined by its message update rules, initialization procedure, and stopping conditions, which together govern the evolution of beliefs across the graph during decoding.

\emph{Messages:} The check-to-error messages at iteration $t$ are:
\begin{equation}
\mu_{i \rightarrow j}(t) =   \kappa_{i,j}(t)\, (-1)^{\sigma_i} 
\min_{j' \in \mathcal{N}(i) \setminus \{ j \}}\left|
\nu_{j' \rightarrow i}(t-1)
\right|,
\label{eq:cntovn}
\end{equation}
\noindent where $\displaystyle \kappa_{i,j}(t) =  \mathrm{sgn}\bigg\{\prod_{j' \in \mathcal{N}(i) \setminus \{ j \}} 
\nu_{j' \rightarrow i}(t-1)\biggr\}$.

\noindent The error-to-check messages are then computed via:
\begin{equation}
    \nu_{j \rightarrow i}(t) =  {\displaystyle
    \Lambda_j(t) + \sum\limits_{i' \in \mathcal{N}(j) \setminus \{i\}} \mu_{i' \rightarrow j}(t).}
    \label{eq:vntocn}
\end{equation}

\emph{Initialization and stopping:}
We initialize by setting the initial beliefs and biases as the log-likelihoods of the error priors $\Lambda_j(0)=\nu_{j\rightarrow i}(0) =\log\frac{1-p_j}{p_j}$ and set the initial marginals from user input $M_j(0)=M_j$. 
After each iteration of message passing, we compute the new marginal $M_j(t)$ and the hard decision $\hat{e}_j(t)$ for each error node $j$:
\begin{eqnarray}
    M_j(t) &=& {\displaystyle \Lambda_j(t) + \sum_{i'\in\mathcal{N}(j)}\mu_{i'\rightarrow j}(t),}
    \label{eq:posteriors} \\
       \hat{e}_j(t) &=&  {\displaystyle \text{HD} \bigl( M_j(t) \bigr), ~~ \text{for}~\text{HD}(x) = \tfrac{1}{2}\bigr(1-\text{sgn}(x)\bigr).} \nonumber
\end{eqnarray}

If the obtained error estimate $\hat{e}_j(t)$ satisfies the parity-check equation
$\mathbf{H}\bm{\hat{e}}(t)=\bm{\sigma}$, the algorithm is considered to have converged and the error vector $\hat{e}_j(t)$ is returned. 
Otherwise we move on to the next iteration until a maximum number of iterations $t = T$ has been reached, in which case DMem-BP is deemed unsuccessful.

\textit{Bias term:}
In DMem-BP the biases are updated via the equation
\begin{equation}
    \Lambda_j(t)  = (1-\gamma_j) \Lambda_j(0) + \gamma_j M_j(t-1).
    \label{eq:relay_update}
\end{equation}
We use $\bm{\Gamma} = \{\gamma_j\}_{j\in[N]}$ to denote real-numbers specifying the memory strength for each error node $j$. DMem-BP therefore consists of a flexible family of decoders parameterized by the choice of memory strengths $\bm{\Gamma}$. The special case of all $\gamma_j = 0$ reduces to the standard BP algorithm.

\paragraph{Relay-BP-$S$} We combine DMem-BP instances sequentially into a relay ensemble known as Relay-BP. Each DMem-BP instance is be called a \textit{leg} of the relay. To describe the relay ensemble properly we require some initial setup. For a given decoding problem $(\mathbf{H}, \mathbf{A}, \mathbf{p})$ the Relay-BP-$S$ algorithm is fully specified by the: 
\begin{itemize}[noitemsep]
    \item number of solutions sought $S$,
    \item maximum number of relay legs $R$, 
    \item maximum number of iterations for each leg $T_r$,
    \item memory strengths for each leg $\bm{\Gamma}_r = \{\gamma_j(r)\}_{j\in[N]}$,
\end{itemize}
\noindent 
for $r \in \{1, ... R\}$. 
The first leg of Relay-BP-$S$ applies DMem-BP initialized with marginals and memory strengths $M_j(0) = \log\frac{1-p_j}{p_j}$ and $\bm{\Gamma}_0= \{\gamma_j(0)\}_{j\in[N]}$. 
The $r^{th}$ leg's marginals are initialized with the $(r-1)^{th}$ leg's final marginals, thereby passing information forward through the relay. Each leg stops upon finding a solution or reaching the iteration limit $T_r$. The algorithm stops when $R$ legs have executed or $S$ solutions have been found, and returns the lowest-weight solution found among all legs, where solution $\mathbf{\hat{e}}$ has weight $w(\mathbf{\hat{e}}) = \sum_j  \hat{e}_j \log\frac{1-p_j}{p_j}$. Pseudocode for Relay-BP is shown in Listing~\ref{alg:Relay} in the appendix, and a reference software implementation is available~\footnote{\url{ https://github.com/trmue/relay}}.

\section{Realization of Relay-BP Hardware Decoder}

\subsection{Decoder Architecture}

\begin{figure}[htb]
  \centering
  \includegraphics[width=0.95\linewidth]{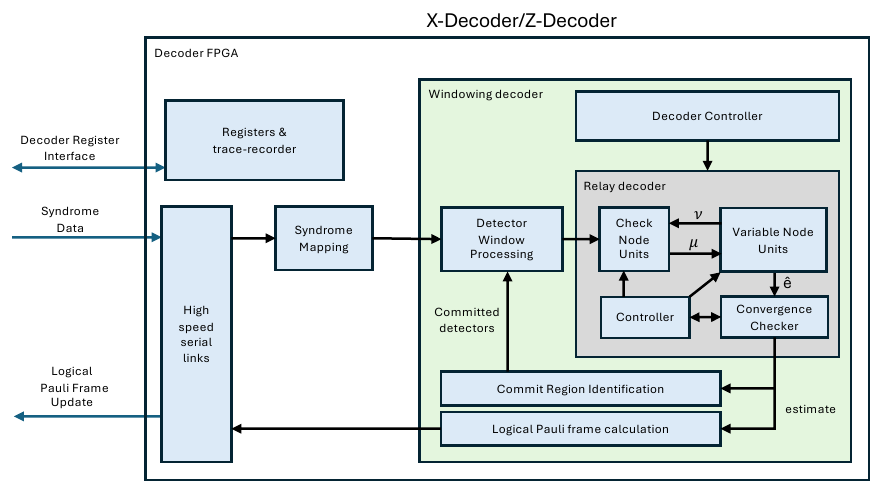}
  \caption{Overview of the Relay-BP FPGA decoder architecture. An external \textit{nest} handles configuration control and includes adapters to conform the I/O data streams to the requirements of the decoder. The windowing decoder creates a window of detectors from the input data stream of syndromes and converts the output error vector $\mathbf{\hat{e}}$ into an update to the logical Pauli frame via the logical action matrix $\mathbf{A}$. The central Relay decoder handles execution of the Relay-BP algorithm via a collection of specialized compute nodes the correspond 1:1 with the decoding graph $G$. The connectivity of $G$ is translated into FPGA wiring defining the mapping of messages between the compute nodes.}
  \label{fig:decoder-architecture}
\end{figure}

\begin{figure}[h!]
    \centering
    \includegraphics[width=0.8\linewidth]{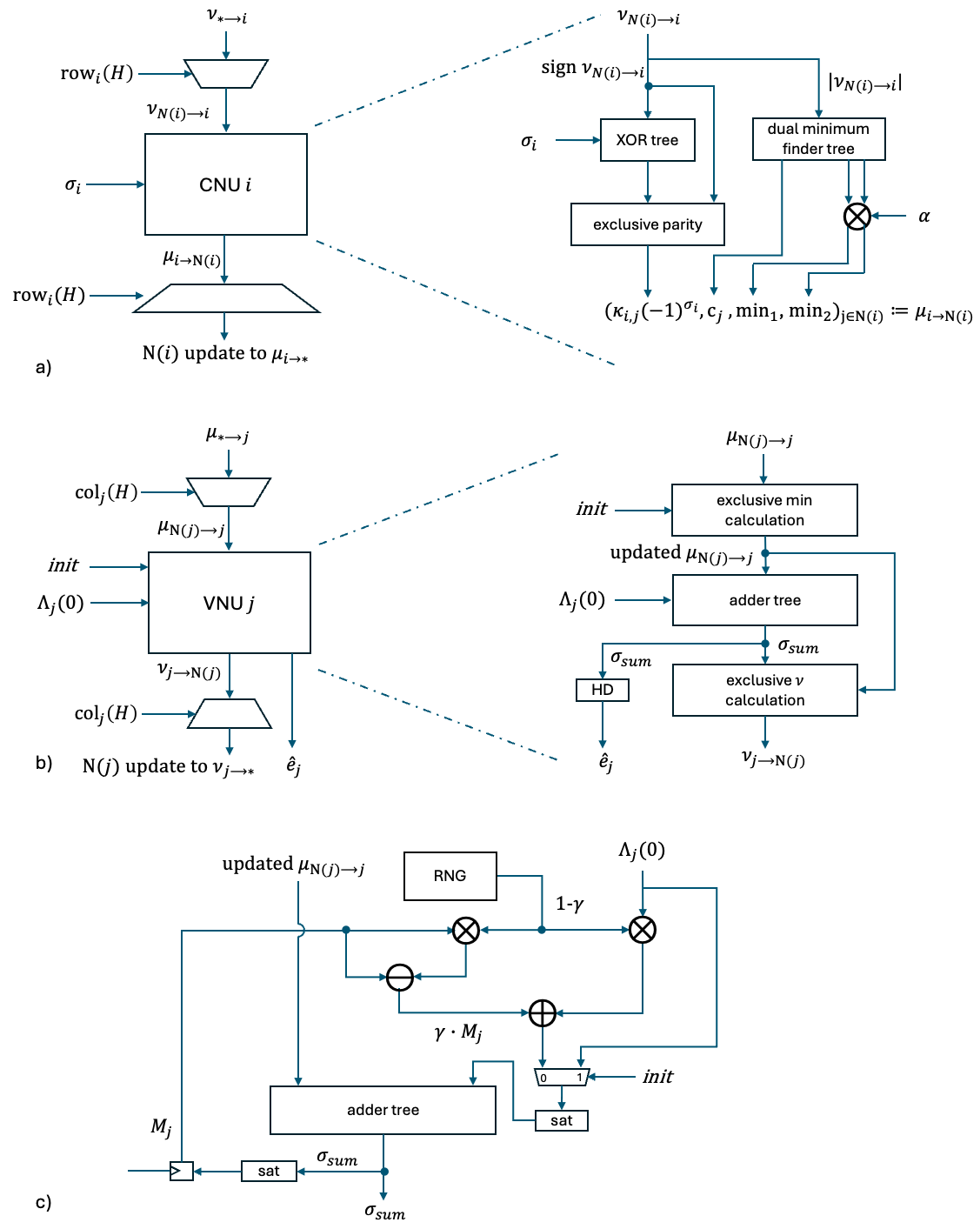}
    \caption{a) Schematic of the check node unit (CNU). We note that the exclusive minimum is not fully calculated in the CNU. Instead only the minimum and second minimum are calculated in the CNU with the exclusive minimum being calculated in the VNU at a later time. b) Schematic of the variable node unit (VNU). c) Schematic of the Relay adder that is used in the VNUs.}
    \label{fig:logic-blocks}
\end{figure}

An overview of the decoder architecture is shown in Fig.~\ref{fig:decoder-architecture}. At the top level, the decoder receives syndrome data and outputs an update to the logical Pauli frame. The input path includes a mapping stage to match the index order of the the input data stream to the decoder configuration. A detector window processing component calculates parities of groups of measurements~\cite{stim}, which is a pre-processing step that allows the decoding problem to be independent of a gauge degree of freedom. The architecture supports windowed decoding~\cite{Iyengar2012, Skoric2022} by identifying the \textit{commit region} in a decoding window and processing of subsequent windows incorporating non-committed detectors from prior windows. The selected detector window is passed into the core Relay decoder component.

More specifically, we implement a (W,C)-sliding window decoder, where the window size $W$ is statically fixed to the (bare) code distance $d$ (\textit{i.e.} $W=10$ for the Loon code and $W=12$ for the gross code), while the commit width $C$ remains configurable at runtime. For each decoding configuration --- X, Z, or XYZ --- the window processing is realized with a handshaking pipeline that supports back-pressure. See Algorithm~\ref{alg:sliding-window-decoding} in the appendix for further details.

% Detectors are derived either directly from syndrome measurements or as effective detectors based on the final codeword measurement. At the conclusion of a memory experiment, which is a repeated sequence of syndrome cycles, a final measurement is performed in the X or Z basis on the code qubits, yielding a (potentially invalid) codeword. This codeword is multiplied by the bare decoding matrix $H_\mathrm{bare}$ \blake{we need to define H\_bare} to compute the final syndrome $\sigma_\mathrm{final}$.

Given that Relay-BP is a message passing algorithm, the central concern governing the implementation is how to arrange for multiple compute resources to access the message data with minimal contention. Building upon the strategy described by Valls et al.~\cite{Valls-et-al-2021}, we assign a unique computational resource to every variable node and check node within the decoding graph, and use FPGA wiring to fix the interconnect between these compute units. This trivially solves the contention issue by placing message data in the wiring fabric rather than a centralized memory. Our design has Variable Node Units (VNUs) and Check Node Units (CNUs) that compute their corresponding outputs in a single FPGA clock cycle, so that two clock cycles are sufficient for a complete BP iteration in a flooding message schedule. A controller initializes these units with appropriate priors and is signaled by a convergence checker when each DMem-BP leg in the Relay-BP ensemble has found a solution. For what follows we will drop the variable $t$ to make things simpler.

\paragraph{Check Node}

Figure~\ref{fig:logic-blocks}a illustrates the block design of the Check Node Unit (CNU). Each CNU processes a single row $i$ of the decoding matrix, denoted $\mathrm{row}_i(\mathbf{H})$. It takes two inputs:
\begin{itemize}
\item A vector of incoming messages $(\nu_{j \rightarrow i})_{j \in \mathcal{N}(i)}$ from neighboring variable nodes, labeled collectively as $\nu_{\mathcal{N}(i) \rightarrow i}$.
\item The value of the local parity check (or detector), $\sigma_i$.
\end{itemize}
From these inputs, the CNU computes a vector of outgoing messages $\mu_{i \rightarrow \mathcal{N}(i)}$, defined here as  the tuple $(\kappa_{i,j}(-1)^{\sigma_i}, c_j, \mathrm{min}_1, \mathrm{min}_2)_{j \in \mathcal{N}(i)}$, which forms a subset of the full message set $\mu_{i \rightarrow *}$. The CNU operates on a message representation that separates sign and magnitude components, i.e., $\nu = (\nu_s, \nu_v)$.

Importantly, the exclusive minimum and exclusive product required by equation~\eqref{eq:cntovn} are not computed individually for each $i$,$j$ pair. Instead, the non-exclusive minimum and product are computed once per fixed $i$, as they remain constant across the neighborhood. The exclusive product is then derived from these constants. For efficiency, the exclusive minimum is only partially computed within the CNU and is passed to the Variable Node Unit (VNU) for final resolution. This optimization enables synthesis tools to further streamline the design.

To support this deferred computation, a selector value $c_j$ is included in the output, allowing the receiving VNU to determine which of the two smallest elements corresponds to the appropriate exclusive minimum for node $j$. The complete output message is thus a vector of tuples of the form $\mu = (\mu_s, \mu_c, \mu_{\mathrm{min}_1}, \mu_{\mathrm{min}_2})$.

Additionally, the value path incorporates a min-sum belief propagation scaling factor $\alpha$, empirically shown to improve convergence. We adopt $\alpha = 1 - 2^{-t}$, where $t \in \mathbb{N}$ is the current iteration index~\cite{6940497}. In practice, our hardware description language (HDL) implementation of the CNU also depends on $\mathrm{row}_i(\mathbf{H})$, and we rely on synthesis tools to convert multiplexers on input and output message vectors into optimized FPGA wiring.

\paragraph{Variable Node}

Detail of the VNU design is shown in Fig.~\ref{fig:logic-blocks}b and ~\ref{fig:logic-blocks}c.
A VNU processes one column $j$ of the decoding matrix that we denote $\mathrm{row}_j(\mathbf{H})$, implementing both Eqns.~\ref{eq:vntocn} and~\ref{eq:posteriors}. It takes three inputs: 
\begin{itemize}
  \item A vector of incoming messages $(\mu_{i \rightarrow j})_{i \in \mathcal{N}(j)}$ from neighboring variable nodes, labeled collectively as $\mu_{\mathcal{N}(j) \rightarrow j}$;
  \item The error priors $\Lambda_j(0)$; and
  \item An initialization signal, \textbf{init}, used by the controller to mark the first Relay-BP iteration.
\end{itemize}

\noindent The VNU then generates a hard error decision vector $\hat{e}_j$, and a vector of output messages $\nu_{j\rightarrow \mathcal{N}(j)}$ which is a subset of $\nu_{j\rightarrow *}$. For standard BP, the VNU design of Fig.~\ref{fig:logic-blocks}b is sufficient and follows directly from Ref.~\cite{Valls-et-al-2021}. To implement DMem-BP, we replace the adder tree of Fig.~\ref{fig:logic-blocks}b with the design of Fig.~\ref{fig:logic-blocks}c, which adds a local pseudo-random number generator (RNG) to randomize the memory strength of each node as well as a bias term that is updated on each BP iteration. As with the CNU, we rely upon the synthesis tool to convert the muxes on the input and output vectors into wiring.

As with the CNU, the VNU and the adder detailed in Figure~\ref{fig:logic-blocks}c do not calculate the Relay-BP messages exactly as defined in equations~\eqref{eq:cntovn} and \eqref{eq:vntocn}. This is due to various implementation optimizations. As such the margins $M_j$ in Figure~\ref{fig:logic-blocks}c are actually the result of the full marginals after passing through a Saturation/clipping process (SAT). Here $\sigma_{sum}$ is the closest representation of the full marginals.

The execution of the Relay-BP algorithm emerges by the way in which the controller configures the VNU, passing priors from one leg to the next, and configuration of the RNG to scramble the memory strengths for each leg.

\subsection{Reduced-Precision}
Standard Belief Propagation is defined as real-valued in the log-likelihood domain. Typical implementations on standard computing architectures implement BP using IEEE floating-point values. Our empirical studies show that 16-, 32-, and 64-bit floating point representations achieve nearly identical logical error rates and BP iteration counts when applied to qLDPC decoding. We may further reduce gateware area using fixed-point values, where \texttt{fpM.N} would indicate $M$ bits being employed to represent the integer part of the values and $N$ bits being employed to represent the fractional part of value. However, without loss of generality, one may recast BP to use purely integer values by incorporating scale factors.

We define \texttt{intN.S.M} to indicate $N\in\mathbb{N}$ bits employed to represent the integer value with a scale factor $S\in\mathbb{R}^+$ to translate log-probability domain real values into the integer domain, and a scale factor $M\in\mathbb{R}^+$ related to the memory strength multiplication. 

Applying a scale factor $S$ is a convenient choice as BP is isomorphic to this. Specifically, the following BP variables scale under $S$: priors $\Lambda_j(0)$, messages $\mu_{i\rightarrow j}$ and $\nu_{j\rightarrow i}$, and posteriors $M_j$. Memory strengths $\gamma\in\mathbb{R}$ for DMem-BP or Relay may be treated in a similar fashion, where their multiplication with any given rescaled log-probability domain value $\mu_\textrm{int}$ is translated as follows:
$$\mu_\textrm{int}\cdot\gamma \rightarrow \round{\mu_\textrm{int}\gamma_\textrm{int} / M}, \quad \gamma_\textrm{int} := \round{\gamma\cdot M}$$
where we employ standard rounding of fractional values to integers to minimize quantization errors.

\subsubsection{Practical Considerations}
Our hardware implementation employs an explicit sign bit for messages $\mu_{i\rightarrow j}$ and $\nu_{j\rightarrow i}$, while we use twos' complement values for posterior values $M_j$, and unsigned values for priors $\Lambda_j(0)$.

We make the gateware-area efficient choice to restrict the memory-strength scaling factor $M$ to powers of 2, \textit{i.e.} $M=2^m$ for some $m \in \mathbb{N}$, such that the division-and-round by $M$ may be implemented via a simple unsigned\footnote{Given the choice of employing an explicit sign bit for log-probability domain values.} bitwise shift-right. This is equivalent to defining the number of fractional bits (as in \texttt{fp0.m}) to be accounted for during multiplication. Since practical values for the memory strengths range $\gamma\in[-1,1]$ we instead employ $\beta = (1 - \gamma) \in [0,2]$ to render $\beta$ positive and instead employ:
$$\mu_\textrm{int}\cdot\gamma \rightarrow \round{\mu_\textrm{int}(M-\beta_\textrm{int}) / M}, \quad 
\beta_\textrm{int} := \round{\beta\cdot M}$$

For example, given a Relay decoding memory strength interval of $\gamma\in [-0.24,0.66]$ and a memory-strength scaling factor $M=8$ we work with $\beta\in [0.34,1.24]$ or $\beta_\textrm{int}\in [3,10]$.

We further reduce the logic requirements by simplifying the multiplication: Instead of implementing a full multiplier for $\mu_\textrm{int}\beta_\textrm{int}$, we expand each bit of the bitwise representation of $\mu_\textrm{int}$ to $\beta_\textrm{int}$, shift right by $m$ places, then null all effective factional bits before summing resulting values for the total result. See Table~\ref{table:low-logic-memory-strength-multiplication} in the appendix for a practical example.

\begin{figure}[htbp]
    \centering
    \includegraphics[width=0.99\linewidth]{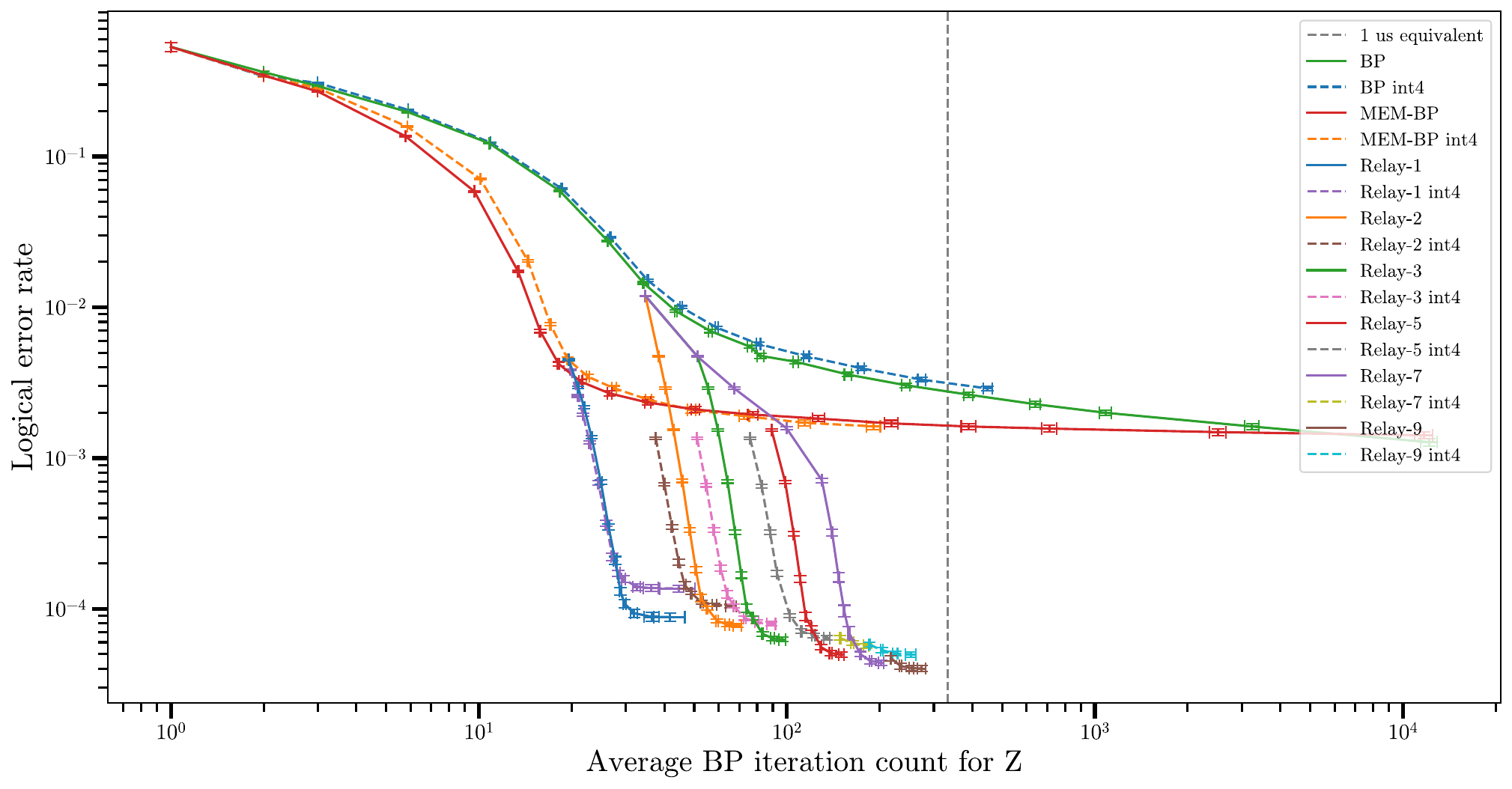}
    \caption{Contour plots along the range of maximum-iteration counts for BP and DMem-BP or maximum-leg counts $R$ for Relay of logical error rate vs average BP iteration count for decoding Z syndromes of a 12 syndrome cycle and a final noiseless cycle of the gross code. We show results from standard BP as well as DMem-BP ($\gamma_0=0.1$) and Relay-S ($\gamma_0=0.1$, $\gamma\in[-0.24,0.66]$, $\beta_\textrm{int}\in[3,10]$, $T_0=T_r=60$), for a software implementation of Belief Propagation in float-64, as well as software emulation of the FPGA logic at \texttt{int4.2.8} precision. For Relay-BP, we observe an improvement in the average number of iterations to converge for the integer implementation at a small cost to logical error rate.
    \label{fig:cutoff-plot-integer}}
\end{figure}

We compare the impact of the taken approximations on BP computation for standard BP, DMem-BP and Relay-BP-S in Figure~\ref{fig:cutoff-plot-integer}: For the depicted integer precision \texttt{int4.2.8} the qualitative decoding runtime characteristics remain unaffected.

% these sections removed for first version
% \input{hardware-details}

\section{Resource utilization}

\begin{figure}[htb]
\centering
\begin{minipage}[t]{0.3\textwidth}
    \vspace{0pt} % Align top
    \centering
    \includegraphics[width=\linewidth]{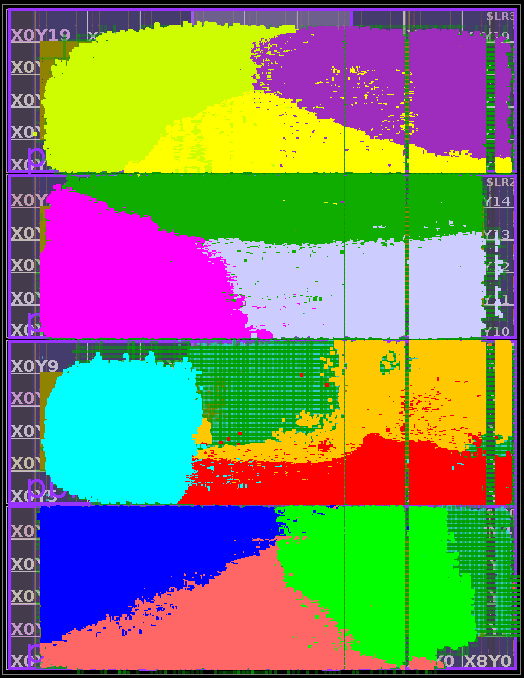}
    % \caption{}
    % \label{fig:floorplan}
\end{minipage}
\hfill
\begin{minipage}[t]{0.48\textwidth}
    \vspace{0pt} % Align top
    \centering
    \begin{tabular}{|l|l|}
    \hline
    \textbf{Decoder} & \textbf{[[144,12,12]]-X} \\
    \hline
    Precision & 4-bit integer + sign \\
    Window width & 12 \\
    Decoder Cycle Time & 12 ns \\
    FPGA Cycle Time & 4 ns  \\
    Flip-Flops & 540,767 (6.62 \%) \\
    LUTs & 2,106,738 (51.56 \%)\\
    LUTRAM & 14,052 (1.47 \%)\\
    BRAM & 29.5 (1.37 \%)\\
    Total Power & 58.025 Watts \\
    PIN Counts &  45 User Pins \\
    \hline
    \end{tabular}
    % \captionof{table}{Decoder Synthesis, Place and Route summaries for a AMD XCVU19P device. The nest and decoder are separated into independent clock domains, so we report cycle times for each.}
    % \label{tab:resources}
\end{minipage}
\caption{(Left) Floorplan of a 12-cycle gross code CSS decoder for X in an AMD XCVU19P FPGA. Colors reflects logic instantiated due to entries of the decoding matrix local to a unique syndrome cycle within the 12-cycle wide decoding matrix. (Right) Resource utilization table after synthesis and place and route. The decoder has it's own independent clock domain, so we report it's cycle times separately.}
\label{fig:resources}
\end{figure}

The implementation strategy of this decoder is resource intensive, requiring large FPGAs to fit the design within a single FPGA. For the size of decoding matrix encountered in a gross code memory experiment, we used AMD's XCVU19P FPGA, which is the second largest FPGA in AMD's product catalog in terms of available resources such as block RAM and look-up tables (LUTs). A summary of the resource utilization of a \texttt{int4} precision gross code decoder is shown in Fig.~\ref{fig:resources}. Notably, the design uses over half of the available LUTs. Since the message data is stored in wiring rather than centralized memories, very little block RAM is used. Strategies to improve the resource footprint of the decoder will be discussed in a future revision of the manuscript.

Despite the large resource utilization, we achieve timing closure of the design even when operating the decoder at a 12ns clock period, which allows an entire Relay-BP iteration to execute in 24ns. We observe similar timing performance over a range of message precisions.

Table \ref{table:resource-breakdown} shows a breakdown of the LUT and flip-flop usage. It becomes obvious that 97\% of the LUTs are consumed in the BP decoder, where the VNUs and CNUs dominate with 67\% and 23\%, respectively. 

\begin{table}[ht]
\centering
\begin{tabular}{|l|l|l|}
\hline
\textbf{Component} & \textbf{LUTs} & \textbf{Flip-Flops} \\
\hline
VNUs & 67.2\% & 36.2\% \\
CNUs & 22.6\% & 13.1\% \\
FSM & 0.1\% & 0.4\% \\
BP decoder other & 7.1\% & 41.3\% \\
windowing decoder & 1.0\% & 2.3\% \\
HSS links & 1.3\% & 2.9\% \\
other & 0.8\% & 3.7\% \\
\hline
\end{tabular}
\caption{LUTs and flip-flops breadkown by component}
\label{table:resource-breakdown}
\end{table}

\section{Verification}

\begin{figure}[htb]
  \centering
  \includegraphics[width=0.95\linewidth]{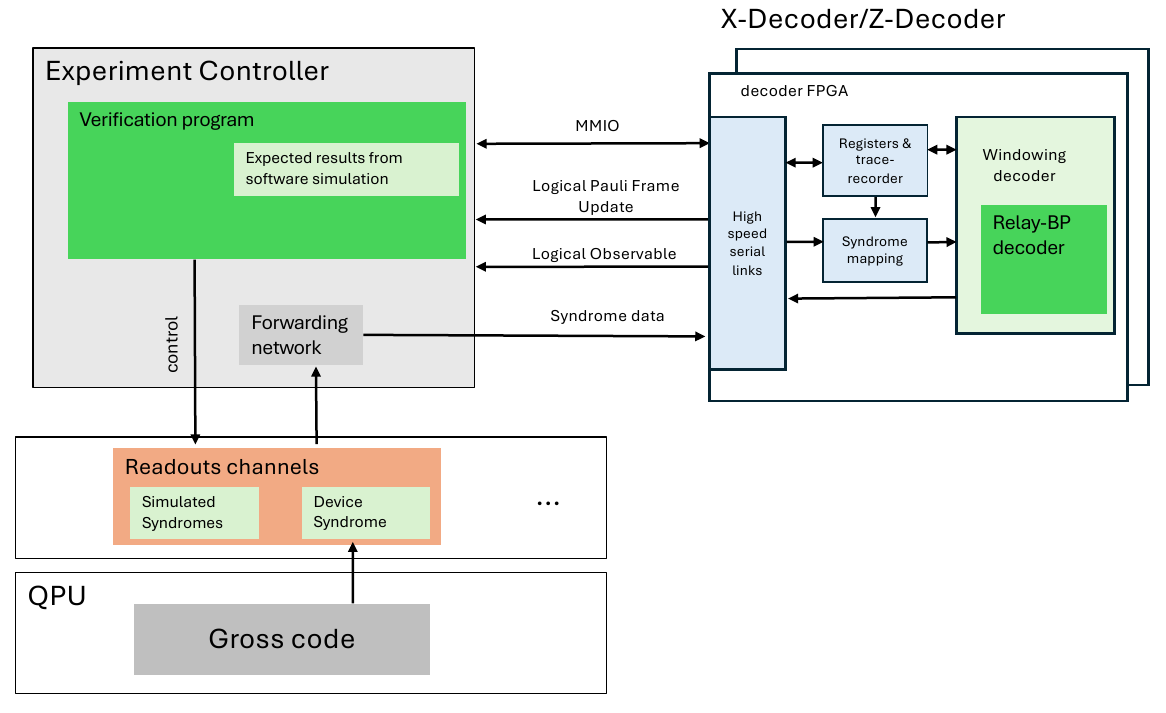}
  \caption{Verification test hardness for the FPGA decoder. A real-time controller is connected to the decoder FPGA and to readout data streams from the quantum control system. Expected vs actual outputs are compared for a variety of synthetic syndrome data streams. The same harness can also forward actual measurement data if an appropriate quantum processor (QPU) is attached.}
  \label{fig:decoder-verficiation-setup}
\end{figure}

\subsection{QEC hardware verification setup}

To validate the performance of the decoder and correct routing of the data, we prepare a test harness that mimics integration into a larger control system, but allows injecting synthetic syndrome data streams. This is done in addition to the analysis of the hardware simulation results, which already showed the feasibility of the implementation. The experimental setup comprises a real-time capable experiment controller which executes a verification program and feeds synthetic syndrome measurement data to the decoder FPGA via the high-speed serial links shown in Fig.~\ref{fig:decoder-verficiation-setup}. Our test harness may be configured to forward qubit measurement data to the decoder when integrated into IBM's quantum control system.

The approach to validate the Relay-BP and the surrounding sliding window decoder is to make sure that hardware results match exactly the software emulation model. 

\subsection{Relay-BP decoder verification}

The Relay-BP decoder is embedded in the sliding window decoder shown in Fig.~\ref{fig:decoder-architecture}. For a window size of W, W syndrome cycles of data are provided to the decoder using the decoder register interface. After the decoding process finishes, we check convergence, number of BP iterations, and equivalence of the estimated error vector $\mathbf{\hat{e}}$ to that produced by software simulation. The correctness of this verification is the basis for the following sliding window decoder verification step.

\subsection{Sliding window decoder verification}

The sliding window decoder is used to integrate the decoding into the system. Before using it, mappings of readout channels to syndrome and code-word bits are configured in the syndrome-mapping table. For the relay decoder the maximum number of iterations for the initial DMem-BP leg, the number of relay legs, and the maximum iterations for those legs are configured to match the software simulation settings. The readout channels are primed with the syndrome bits derived from a noise model simulation of the syndrome cycle circuit.
Now the experiment controller can start the verification. The readout channels will produce syndrome readouts every 1 $\mu$s, which are routed via the high speed serial connection to the decoder FPGA. Once all syndrome bits are collected, the syndrome is put into a FIFO and once the syndrome FIFO contains enough syndromes to invoke the Relay-BP decoder (window cycle/commit cycle), the decode process is started. When the process completes, results are stored within the sliding window decoder. This is iterated when new syndromes arrive at the decoder.

As soon as the experiment controller instructs the readout channels to generate the codeword, it is routed in the same fashion to the sliding window decoder which is calculating the logical qubit vector. This is sent back to the logical qubit storage where it can be used by the verification program to match it against the expected results from the software simulation. The process checks if the window decoding finished with valid results, the number of window decoder invocations, the correctness of the Pauli frame, the total number of BP decoder iterations, and how many sliding window decoder invocations converged.

To get precise timing information for the individual steps the decoder FPGA provides hardware trace information which allows observation of when the decoders start and stop, when the syndromes and code-words arrived, and when the results like the logical Pauli frame are produced.

\section{Results}

\begin{figure}[htbp]
  \centering
  \begin{subfigure}[b]{0.49\textwidth}
    \centering
    \includegraphics[width=\linewidth]{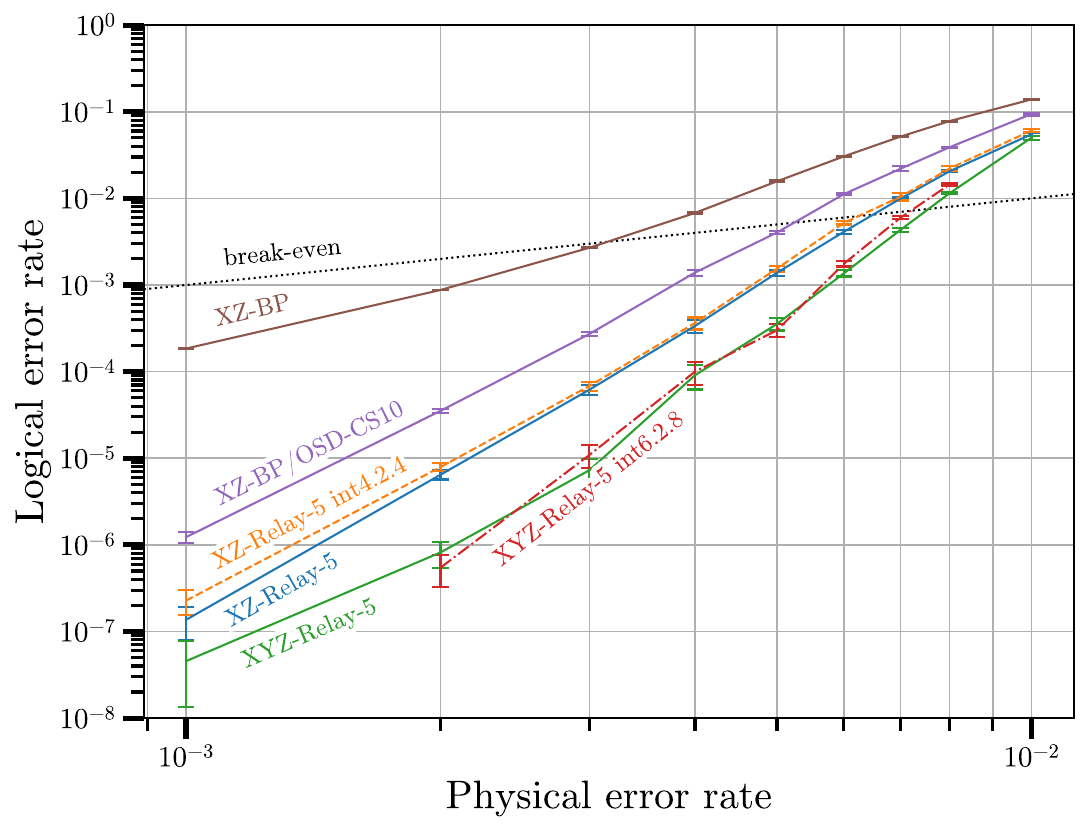}
    \caption{Logical error rate for various physical error rates
     in the context of global decoding over 10+1 cycles for the Loon memory experiment.}
  \end{subfigure}
  \hfill
  \begin{subfigure}[b]{0.49\textwidth}
    \centering
    \includegraphics[width=\linewidth]{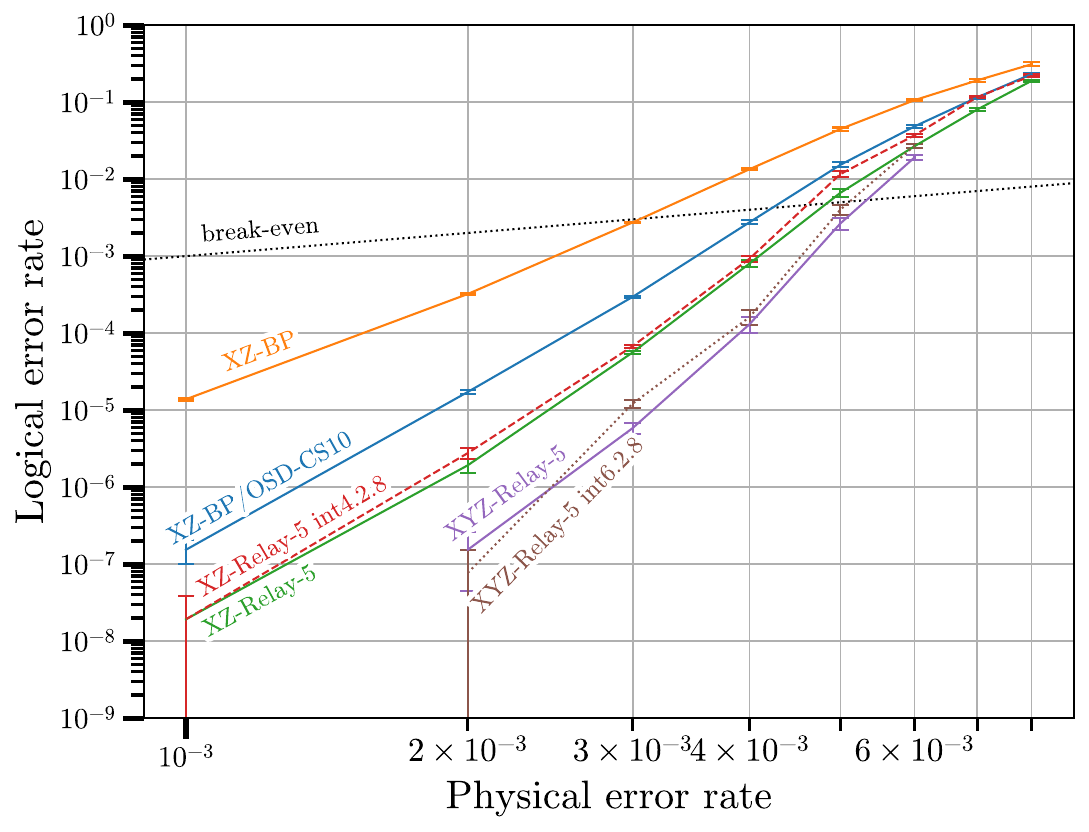}
    \caption{Logical error rate for various physical error rates
     in the context of global decoding over 12+1 cycles for the gross code memory experiment.}
  \end{subfigure}

  \caption{Decoding performance of the integer-based Relay-N FPGA implementation for the Loon and Kookaburra memory experiments. For both the smaller code on Loon, and the gross code on Kookaburra, int4 precision is sufficient to match the accuracy of floating point in the case of split XZ decoding. Correlated XYZ decoding requires int6 precision to observe equivalent accuracy to floating point. Parameters employed: DMem-BP $\gamma_0=0.125$, $T=12800$; Relay-S $\gamma_0=0.125$, $\gamma\in[-0.24,0.66]$, $\beta_\textrm{int}\in[3,10]$, $T_0=80$, $T_r=60$, $R=600$.}
   \label{fig:error-rate}

\end{figure}

Summary results of the accuracy of the FPGA decoder applied to both $[[56, 2, 10]]$ and $[[144, 12, 12]]$ memory experiments are shown in Fig.~\ref{fig:error-rate}. The reduced precision design achieves parity in logical error rate with a floating-point software implementation for a suitable choice of integer format. Data in decoding performance plots has been generated using existing software implementations of standard Belief Propagation, DMem-BP and Relay-BP ~\footnote{\url{ https://github.com/trmue/relay}}, as well as a custom software implementation of the same with the specific integer arithmetic and windowing method employed with the FPGA, also acting as the reference implementation for cross-validation of the FPGA implementation.

This design achieves high throughput and constant per-iteration latency, making it suitable for real-time quantum error correction. In particular, timing data from a split X/Z Relay-BP decoder of the gross code memory experiment completes a BP iteration in 24ns. Additional data (appendix Fig.~\ref{fig:decoding-runtime}) explore contours of BP iterations required for various logical error rates. At the lowest examined physical error rate of $p = 10^{-3}$, we observe that the average number of iterations required for BP to converge is roughly 20. This implies that a full 12-cycle window may be decoded in 480ns, giving an average per cycle decoding time of 40ns.

 To validate the Relay-BP decoder and its integration into a sliding-window architecture, we implemented a hardware verification setup that mimics a real control system while allowing injection of synthetic syndrome streams. The test harness includes a real-time experiment controller connected to the decoder FPGA via high-speed serial links, with readout channels configured to emit precomputed syndromes from software simulations. Verification ensures hardware results match software outputs for convergence, iteration counts, and error estimates. The sliding-window decoder is tested under realistic timing constraints, using configurable mappings and iteration limits. Hardware trace data provides precise timing for decoding steps. This approach confirms correct data flow and decoding behavior, establishing readiness for integration with actual QPU hardware.
 
\section{Looking Forward}

We have presented the hardware realization of the Relay-BP decoder on an FPGA with high accuracy and state-of-the art per-cycle decoding times. At its core, the decoder algorithm is implemented through a fully parallelized message-passing scheme, assigning dedicated compute units to each variable and check node in the decoding graph. This eliminates memory contention by embedding message exchange in FPGA wiring. Each Variable Node Unit (VNU) and Check Node Unit (CNU) completes its computation in one clock cycle, enabling a full BP iteration in two cycles under a flooding schedule.

To further reduce resource usage, we replace floating-point arithmetic with scaled integer representations for BP messages and priors. This approach preserves decoding accuracy while significantly reducing gateware area. Experiments confirm that low-bit integer formats (4–6 bits) achieve logical error rates comparable to floating-point BP, enabling efficient and scalable hardware implementations.

Next steps include testing the decoder with real syndrome data, which will stress the Relay-BP algorithm's performance under a different noise model. We also need to extend this work toward decoding logical operations and not just logical memory. Doing so will require solving additional challenges in scaling logic design and will build upon lessons learned from this prototype decoder.

\paragraph{Author Contributions:}  Markus wrote the BP-Decoder gateware; Michael did the Nest/System integration; Thilo wrote the Windowing Decoder, the reference software model, and did the performance plots; Michael/Frank did the validation on hardware; Drew and Blake provided project and strategic direction. All authors contributed to the writing of the paper.

\paragraph{Acknowledgments:} The authors would like to thank Lev Bishop, Thomas Alexander, Andrew Cross, Ted Yoder, Scott Willenborg, Scott Lekuch and team, Kevin Krsulich, and Matthew Walther.

%TC:ignore

\ifthenelse{\boolean{preprint}}{
\bibliography{refs.bib}
}{
\printbibliography
}

%\newpage
\appendix
\begin{center}\textbf{Appendix}\end{center}

\section{Relay-BP Algorithm}

\begin{algorithm}[h]
    \SetKwInOut{Input}{Input}
    \SetKwInOut{Output}{Output}
    \Input{Parity-check matrix $\boldsymbol{\mathrm{H}}$, syndrome $\boldsymbol{\sigma}$, error probabilities $\boldsymbol{\mathrm{p}}$, number of solutions to be found $S$, maximum number of legs of the relay $R$, maximum number of iterations per leg $T_r$, and memory strengths for each leg $\{\boldsymbol{\gamma}(r)\}_{r\in[R]}$.}
    \Output{Solution found, Estimated error $\mathbf{\hat{e}}$}
    $\lambda_j, M_j \left( 0 \right) \leftarrow \log\frac{1-p_j}{p_j}$, $r \leftarrow 0,  s \leftarrow 0, \hat{\mathbf{e}} \leftarrow 	\varnothing, \omega_{\hat{\mathbf{e}}} \leftarrow \infty, \Lambda_j(0) \leftarrow \lambda_j$\;
    \For{$r \leq R$}
      {
        \tcp{Run DMem-BP}
        $\nu_{j\rightarrow i}(0) \leftarrow \lambda_j$\;
        \For{$t \leq T_r$}
          {     
            $\Lambda_j(t) \leftarrow (1-\gamma_j(r)) \Lambda_j(0) + \gamma_j(r) M_j(t-1)$\;
            Compute $\mu_{i \rightarrow j} \left( t \right)$ \tcp*[f]{via Eq.~\eqref{eq:cntovn}}\;
            Compute $\nu_{j \rightarrow i} \left(t \right)$  \tcp*[f]{via Eq.~\eqref{eq:vntocn}}\;
            Compute $M_j \left(t \right)$  \tcp*[f]{via Eq.~\eqref{eq:posteriors}}\;
            $\hat{e}_j(t) \leftarrow \mathrm{HD} \bigl( M_j(t) \bigr)$\;
            \If{$\mathbf{H\hspace{0.0833em}\mathbf{\hat{e}}}(t) = \boldsymbol{\sigma}$}
            {
              \tcp*[h]{BP converged}\;
              $\omega_r \leftarrow w(\mathbf{\hat{e}}) = \sum_j  \hat{e}_j \lambda_j$\;
              $s \leftarrow s + 1$\;
              \If{$\omega_r < \omega_{\hat{\mathbf{e}}}$}
              {
                $\hat{\mathbf{e}} \leftarrow \hat{\mathbf{e}}(t)$\;
                $\omega_{\hat{\mathbf{e}}} \leftarrow \omega_r$\;
              }
              $\boldsymbol{\mathrm{break}}$; \tcp*[f]{Continue to next leg}
            }
            $t \leftarrow t + 1$\;
          }
        \If{$s = S$}
            {
              $\boldsymbol{\mathrm{break}}$; \tcp*[f]{Found enough solutions}
            }
        \tcp{Reuse final marginals for the next leg}
        $M_j(0) \leftarrow M_j \left( t \right)$\;
        $r \leftarrow r + 1$\;
      }
    $\boldsymbol{\mathrm{return}}$ $(s > 0)$, $\mathbf{\hat{e}}$;
     
    \caption{Relay-BP-$S$ decoder for quantum LDPC codes}
    \label{alg:Relay}
\end{algorithm}

\section{Further results of decoding performance and runtimes}

\begin{figure}[h!]
  \begin{subfigure}[b]{0.49\textwidth}
    \centering
    \includegraphics[width=\linewidth]{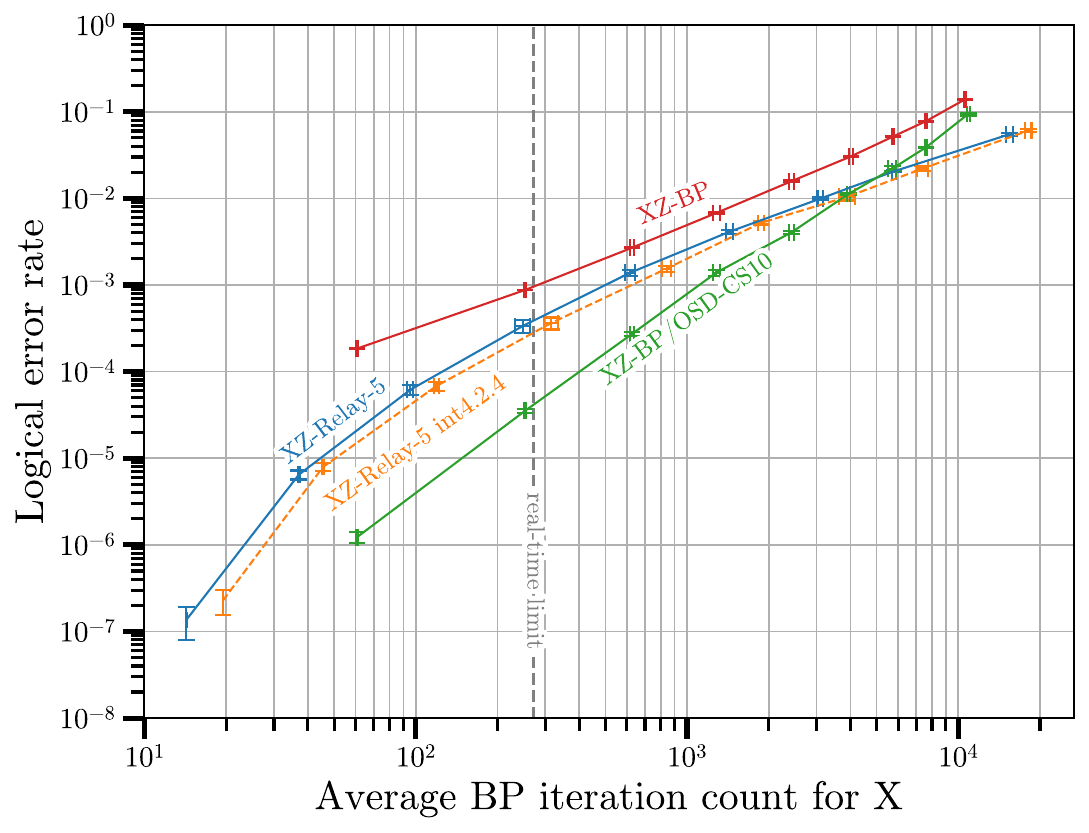}
    \caption{Global decoding over 10+1 cycles for the Loon memory experiment. At a gateware iteration time of 20 ns, a (10,6)-window decoding system permits up to approximately 270 BP iterations per decoding window to sustain real-time decoding.}
  \end{subfigure}
  \hfill
  \begin{subfigure}[b]{0.49\textwidth}
    \centering
    \includegraphics[width=\linewidth]{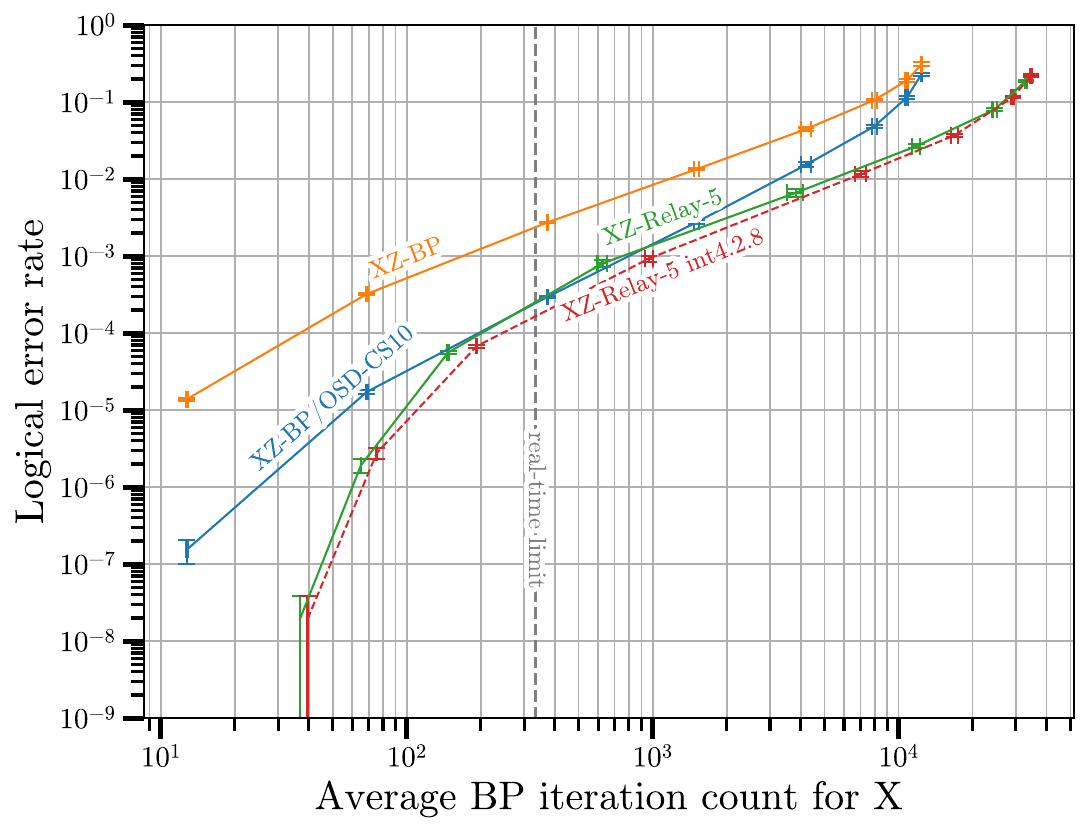}
    \caption{Global decoding over 12+1 cycles for the Kookaburra memory experiment. At a gateware iteration time of 24 ns, a (12,8)-window decoding system permits up to approximately 333 BP iterations per decoding window to sustain real-time decoding.}
  \end{subfigure}

  \caption{Decoding performance of the integer-based Relay-N FPGA implementation for the Loon and Kookaburra memory experiments.
  Contour plots along the range of physical error rates $[0.001,0.01]$ of the average iteration count vs. achieved logical error rates in the context of global decoding. The given real-time budgets assume a (W,C)-sliding window decoder operating over a sequence of IDLE operations spaced at 1 $\mu$s intervals. Parameters employed: DMem-BP $\gamma_0=0.125$, $T=12800$; Relay-S $\gamma_0=0.125$, $\gamma\in[-0.24,0.66]$, $\beta_\textrm{int}\in[3,10]$, $T_0=80$, $T_r=60$, $R=600$.}
   \label{fig:decoding-runtime}

\end{figure}

\section{Reduced logic implementation of the memory strength multiplication}

\begin{table}[h!]
\centering
\begin{tabular}{|c|c|c|c|c|}
\hline
\textbf{Term} & \textbf{Exact} & \textbf{Binary (divided by $2^m$)} & \textbf{Binary (decimals nulled)}& \textbf{Approximation}\\
\hline
$15 \times 7$ & 105 & 0b1101.001 & 0b1011.000 & 88\\
\hline
$8 \times 7$  & 56 & 0b0111.000 & 0b0111.000 & 56\\
$4 \times 7$  & 28 & 0b0011.100 & 0b0011.000 & 24\\
$2 \times 7$  & 14 & 0b0001.110 & 0b0001.000 & 8\\
$1 \times 7$  & 7  & 0b0000.111 & 0b0000.000 & 0\\
\hline
\end{tabular}
\caption{Reduced-logic implementation of memory strength multiplication: floor each term of multiplication by ignoring decimals before summation. In this example we multiply a (scaled) log-probability domain integer value of 15 by a scaled memory strength integer value of 7 at multiplication scale factor of $M=2^3=8$.}
\label{table:low-logic-memory-strength-multiplication}
\end{table}

\section{Sliding-window decoding algorithm}

\begin{algorithm}[h!]
\caption{Sliding-Window Decoding with Commit Region and Final Codeword}
\label{alg:sliding-window-decoding}
\DontPrintSemicolon
\SetKwInput{KwIn}{Input}
\SetKwInput{KwOut}{Output}
\SetKwInput{KwVar}{Variables}
\SetKwInput{KwDesc}{Description}
\KwDesc{Enqueue incoming detectors $\mathbf{d}$ for a full decoding window, also accounting for the carried detector correction $\mathbf{u}$. Decode the window, and update logical Pauli frame $\mathbf{f}$ and to be carried detector corrections before advancing to the next window.}

\KwIn{\;
  Window width $W$ (cycles), commit width $C$ (cycles) with $1 \le C < W$\;
  Detectors-per-cycle $M$, errors-per-window $\tilde{N}$, data qubits $n_q$\;
  Noiseless decoding matrix $\mathbf{\hat H} \in \mathbb{F}_2^{M \times n_q}$\;
  Logical readout matrix $\mathbf{L} \in \mathbb{F}_2^{K \times n_q}$\;
  Windowed parity-check matrix $\mathbf{\tilde{H}}\in \mathbb{F}_2^{WM\times \tilde{N}}$\;
  Windowed logical action matrix $\mathbf{\tilde{A}}\in \mathbb{F}_2^{K\times \tilde{N}}$\;
  Commit mask $m_{\mathrm{com}} \in \mathbb{F}_2^{\tilde{N}}$;  Convergence mask $m_{\mathrm{cvg}} \in \mathbb{F}_2^{WM}$\;
  Inner decoder $\textsc{Decode}(\tilde H, m,\,\cdot)$ returning errors $\mathbf{\hat e} \in \mathbb{F}_2^{\tilde{N}}$
}
\KwOut{\;
  Accumulated logical Pauli frame $\mathbf{f} \in \mathbb{F}_2^K$; Corrected observables $\mathbf{o}_{\mathrm{corr}} \in \mathbb{F}_2^K$\;
}
\KwVar{\;
  Syndrome-cycle $t$, total cycles $T$\;
  Code-qubit codeword $\mathbf{c} \in \mathbb{F}_2^{n_q}$; Check-qubit detectors $\mathbf{d} \in \mathbb{F}_2^{M}$\;
  Logical Pauli frame $\mathbf{f} \in \mathbb{F}_2^{K}$; Current syndrome $\bm{\sigma} \in \mathbb{F}_2^{M}$\;
  Carried detector correction $\mathbf{u} \in \mathbb{F}_2^{M}$\;
  Detector history $\mathbf{D} \in \mathbb{F}_2^{TM}$ (one block of $M$ per cycle $t<T$ )\;
  Window of detectors $\mathbf{\tilde d} \in \mathbb{F}_2^{WM}$\;
}
\BlankLine
$\mathbf{f} \gets \mathbf{0}_K$; $\mathbf{u} \gets \mathbf{s} \gets \mathbf{0}_M$; $t \gets T \gets 0$\;
\Repeat{$t >= T$}{
  \If{\(\textsc{Incoming}\) $\mathbf{c}$}{
    $\bm{\sigma}_{\mathrm{final}} \gets \mathbf{\hat H} \cdot \mathbf{c}$ \tcp*{Final "noiseless" syndrome}
    $\mathbf{d} \gets \bm{\sigma} \oplus \bm{\sigma}_{\mathrm{final}}$ \tcp*{syndrome differencing}
  }\ElseIf{\(\textsc{Incoming}\) $\mathbf{d}$}{
    $\mathbf{d} \gets \mathbf{d}$\;
  }\ElseIf{\(\textsc{Incoming}\) $\textrm{END}$}{
    $\mathbf{d} \gets \mathbf{0}_{M}$\;
  }
  $\bm{\sigma} \gets \bm{\sigma} \oplus \mathbf{d}$ \tcp*{update syndrome}
  $\mathbf{D}[T M : (T+1) M] \gets \mathbf{d}$ \tcp*{Append new detectors}
  $T \gets T+1$\;
  \If{\(t+W<T\)}
  {
    \textbf{repeat} \tcp*{Accumulate full window}
  }
 
  $\mathbf{\tilde d}[0: W M] \gets \mathbf{D}[t M : (t+W) M]$ \tcp*{Extract current window of detectors}
  
  $\mathbf{\tilde d}[0:M] \gets \mathbf{\tilde d}[0:M] \oplus \mathbf{u}$\ \tcp*{Apply carry-in to the \emph{first} cycle of the window}

  $\mathbf{\hat e} \gets \textsc{Decode}(\mathbf{\tilde H}, m_{\mathrm{com}}, \mathbf{\tilde d})$\ \tcp*{Inner window decoding}

  $\mathbf{\hat e}_{\mathrm{com}} \gets \mathbf{\hat e} \odot m_{\mathrm{cvg}}$\ \tcp*{Reduce to commit-errors via mask}

  $\Delta \mathbf{f} \gets \mathbf{\tilde A} \cdot \mathbf{e}_{\mathrm{com}}$ \tcp*{Effects of committed errors}
  $\Delta \mathbf{d} \gets \mathbf{\tilde H} \cdot \mathbf{e}_{\mathrm{com}} $

  $\mathbf{u} \gets \Delta \mathbf{d}[C M : (C+1) M]$ \tcp*{Prepare carry for next window}

  $\mathbf{f} \gets \mathbf{f} \oplus \Delta \mathbf{f}$  \tcp*{Accumulate and slide by the commit stride}
  $t \gets t + C$\;
}
$\textbf{o}_{\mathrm{corr}} \gets (\mathbf{L} \cdot \mathbf{c}) \oplus \mathbf{f}$ \tcp*{Final corrected observables}
\Return $(\mathbf{f},\ \mathbf{o}_{\mathrm{corr}})$.
\end{algorithm}

\end{document}